\documentclass[10pt]{iopart}

\usepackage{graphicx}
\usepackage{dcolumn}
\usepackage{bm}
\usepackage[colorlinks=true,
 linkcolor=black,
 anchorcolor=black,
 citecolor=black,
 urlcolor=black
 ]{hyperref}

\usepackage{verbatim}
\usepackage{braket}
\usepackage{siunitx}
\DeclareSIUnit\gauss{G}
\usepackage{makecell}
\usepackage{xcolor}
\usepackage{lineno}

\begin{document}

\title{Continuously tunable single-photon level nonlinearity with Rydberg state wave-function engineering}

\author{Biao Xu$^{1,\ast}$, Gen-Sheng Ye$^{1,\ast}$,  Yue Chang$^{2,3,\ast,\dagger}$, Tao Shi$^{4,5}$ and Lin Li$^{1,6,\ddagger}$}
\address{$^{1} $MOE Key Laboratory of Fundamental Physical Quantities Measurement, Hubei Key Laboratory of Gravitation and Quantum Physics, PGMF, Institute for Quantum Science and Engineering, School of Physics, Huazhong University of Science and Technology, Wuhan 430074, China}
\address{$^{2}$ Beijing Automation Control Equipment Institute, Beijing 100074, China}
\address{$^{3}$ Quantum Technology R$\&$D Center of China Aerospace Science and Industry Corporation, Beijing 100074, China}
\address{$^{4}$ Institute of Theoretical Physics, Chinese Academy of Sciences, P.O. Box 2735, Beijing 100190, China}
\address{$^{5}$ CAS Center for Excellence in Topological Quantum Computation, University of Chinese Academy of Sciences, Beijing 100049, China}
\address{$^{6}$ Wuhan Institute of Quantum Technology, Wuhan 430206, China}
\address{$^{\ast}$ These authors contributed equally: Biao Xu, Gen-Sheng Ye, Yue Chang.}
\ead{$^\dagger$yuechang7@gmail.com}
\ead{$^\ddagger$li\_lin@hust.edu.cn}

\begin{abstract}
\noindent
Extending optical nonlinearity into the extremely weak light regime is at the heart of quantum optics, since it enables the efficient generation of photonic entanglement and implementation of photonic quantum logic gate.
Here, we demonstrate the capability for continuously tunable single-photon level nonlinearity, enabled by precise control of Rydberg interaction over two orders of magnitude, through the use of microwave-assisted wave-function engineering.
To characterize this nonlinearity, light storage and retrieval protocol utilizing Rydberg electromagnetically induced transparency is employed, and the quantum statistics of the retrieved photons are analyzed.
As a first application, we demonstrate our protocol can speed up the preparation of single photons in low-lying Rydberg states by a factor of up to $\sim 40$.
Our work holds the potential to accelerate quantum operations and to improve the circuit depth and connectivity in Rydberg systems, representing a crucial step towards scalable quantum information processing with Rydberg atoms.

\vspace*{1\baselineskip}
\noindent
\textit{Keywords}: Rydberg atoms, single-photon level nonlinearity, Rydberg interaction, Quantum optics
\end{abstract}

\maketitle
\ioptwocol

\section{Introduction}
\noindent
Single photons usually interact weakly with matter and with themselves.
This non-interacting nature of photons protects them from decoherence during propagation and therefore makes them the best messengers for the distribution of quantum information \cite{obrien2009photonic, pan2012multiphoton, wehner2018quantum}.
However, this remarkable feature also brings obstacles to quantum photonic operations that require strong and tunable nonlinearities at the single-photon level \cite{scully1997quantum, obrien2007optical, kok2010introduction}.
Conventionally, single-photon nonlinearity can be induced by projective measurements and post-selections \cite{knill2001scheme, obrien2003demonstration, okamoto2009entanglement}, but only in an inefficient (probabilistic) way.
Recently, progress has been made in mediating the interaction between individual photons using high-lying Rydberg atoms \cite{pritchard2010cooperative, dudin2012strongly, peyronel2012quantum, firstenberg2013attractive, firstenberg2016nonlinear, tiarks2016optical, busche2017contactless, ornelas2021tunable, chen2021two, lee2023quantum} and achieving efficient quantum photonic operations \cite{li2013entanglement, gorniaczyk2014single, tiarks2014single, tresp2016single, tiarks2019photon, ornelas2020demand, xu2021fast, sun2022deterministic, yang2022sequential, shi2022high, magro2023deterministic, ye2023photonic}.

To further advance the development of scalable multi-photon quantum optics, it is imperative to not only achieve single-photon level nonlinearity but also to make it tunable \cite{chang2014quantum}.
This tunability plays a crucial role in quantum information processing as it enables the implementation of robust and high-fidelity photonic quantum logic operations \cite{han2020error} which is significant for the large scale photonic quantum circuits.
Furthermore, it allows for the precise manipulation of the multi-photon interaction Hamiltonian, holding promise for the realization of novel states of matter made of light, such as photonic Wigner crystal, and the study of intriguing dynamics in quantum phase transitions \cite{otterbach2013wigner, clark2020observation}.

In principle, Rydberg atom-mediated photonic nonlinearity can be tuned via the control of atomic interaction strength, such as changing the principal quantum number $n$ to modify the van der Waals (vdW) interactions coefficient $C_6$.
However, the change of interaction through $n$ can not be achieved continuously and dynamically.
Alternatively, experiments that employ static electric \cite{bohlouli2007enhancement} and magnetic \cite{pohl2009cold} fields to realize F\"{o}rster resonance for the enhancement of atomic interaction have been reported.
Furthermore, the microwave-assisted interaction control \cite{tanasittikosol2011microwave, maxwell2013storage, tretyakov2014controlling, sevinccli2014microwave, zhang2020submicrosecond} also offers an alternative approach to enhance interactions.

An effective way to manipulate the interaction between the atoms is to precisely control their wave-functions.
Rydberg-ground state wave-function dressing has enabled strong interaction between dressed ground-state atoms and found applications in quantum state preparation \cite{jau2016entangling}, quantum logic gate \cite{martin2021molmer}, precision metrology \cite{arias2019realization}, and many-body physics \cite{zeiher2016many}.
Here, we develop a novel Rydberg-Rydberg state dressing protocol via microwave-assisted wave-function engineering.
Based on this technique, we achieve the continuous tuning of Rydberg interaction strength over two orders of magnitude.
To characterize the nonlinearity, we perform light storage-and-retrieval protocol using Rydberg electromagnetically induced transparency (EIT), and analyze the quantum statistics of the retrieved photons.

An immediate application of our protocol is to speed up the quantum operations based on Rydberg dephasing mechanisms \cite{bariani2012dephasing, stanojevic2012generating, bariani2012dephasingpra}.
This dephasing mechanism has been utilized in experimental realizations of several photonic quantum operations with low-lying Rydberg states \cite{busche2017contactless, li2022dynamics, ye2023photonic}, which are less susceptible to many decoherence and loss channels, such as long-lived Rydberg contaminants, residual electric fields, and Rydberg-ground states interaction dephasing \cite{baur2014single, gaj2014molecular, goldschmidt2016anomalous}, than their highly excited counterparts.
However, the dephasing mechanism with lower $n$ accompanies much weaker interaction strengths and requires much longer quantum operation times.
Our protocol tackles this problem by improving the interaction strength by orders of magnitude.
As a result, fast photonic quantum state preparation can be achieved using low $n$ Rydberg states.
We would like to emphasize that accelerating the generation of single-photon sources is just one of the numerous potential applications of our protocol and not the sole focus.
We use the change of the second-order intensity correlation function $g^{(2)}(0)$ of the retrieved photons to characterize the performance and effectiveness of our scheme.

\begin{figure*}[t]
  \centering
  \includegraphics[width=\textwidth]{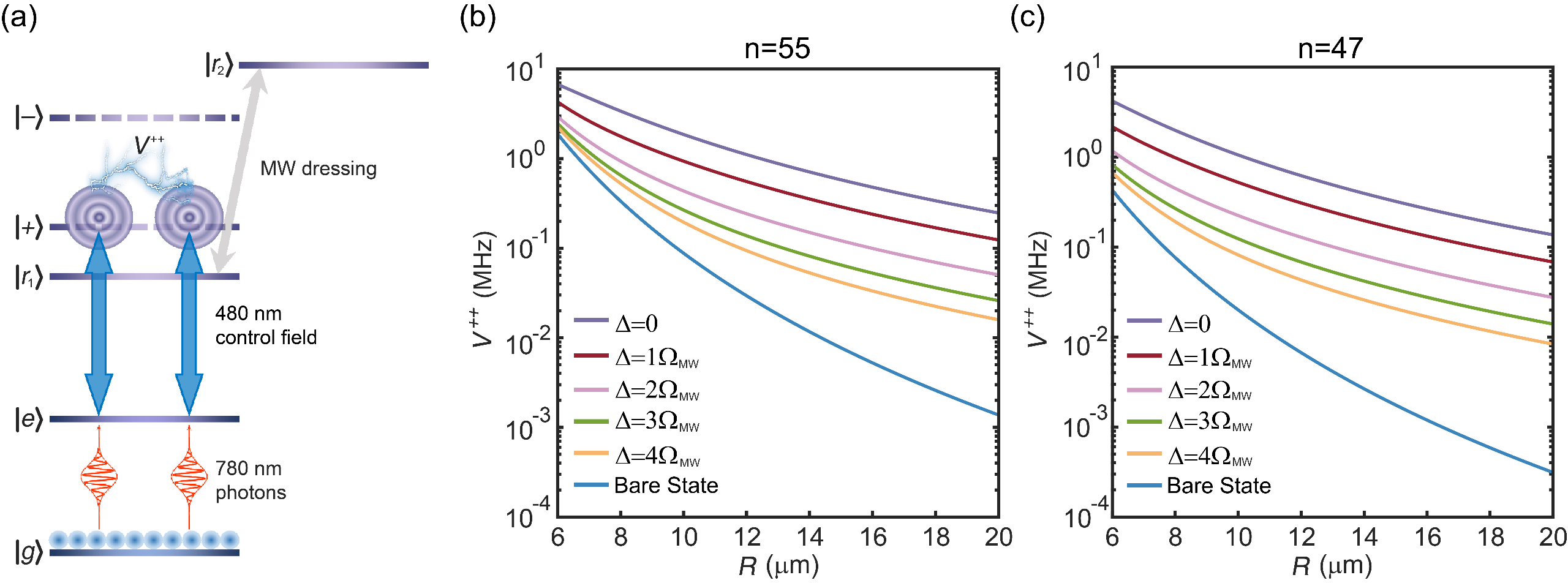}
  \caption{
    Interaction control via Rydberg-Rydberg wave-function dressing.
    (a) The relevant atomic levels: the ground state $\ket{g} = \ket{5S_{1/2}, F=2, m_F=2}$, the intermediate state $\ket{e} = \ket{5P_{3/2}, F=3, m_F=3}$, two Rydberg states $\ket{r_1}$ and $\ket{r_2}$ with different parities, and the MW-dressed Rydberg states $\ket{+}$ and $\ket{-}$ when the microwave is in resonance with two Rydberg states $\ket{r_1}$ and $\ket{r_2}$.
    The \SI{780}{\nano\meter} photons is resonant with $\ket{g} \leftrightarrow \ket{e}$, and the \SI{480}{\nano\meter} control field resonantly couples the $\ket{e}$ and $\ket{+}$ states.}
    (b) The interaction strength $V^{++}$ for the MW-dressed Rydberg states with different $\ket{r_2}$ components, as a function of inter-atomic distances $R$ on a logarithmic scale.
    The six curves are simulated with $\ket{r_1}=\ket{55D_{5/2}, J=5/2, m_J=5/2}$ and $\ket{r_2}=\ket{56P_{3/2}, J=3/2, m_J=3/2}$, with different MW detunings and with MW field turned off (bare state).
    (c) Same as \textbf{b,} but for $\ket{r_1}=\ket{47D_{5/2}, J=5/2, m_J=5/2}$ and $\ket{r_2}=\ket{48P_{3/2}, J=3/2, m_J=3/2}$.
  \label{fig1}
\end{figure*}

\section{Theoretical modeling}
\noindent
Given that the dipole moment $\braket{\hat{d}}$ for a parity-definite state $\ket{r_1}$ is zero, the expectation value for the dipole-dipole interaction operator $\bm{\hat{V_{dd}}}$ on the pair state $\ket{r_1r_1}$ vanishes: $\bra{r_1r_1}\bm{\hat{V_{dd}}}\ket{r_1r_1}=0$, since it scales as $\braket{\hat{d}}^2$.
Therefore, the Rydberg-Rydberg interaction is usually a second-order effect induced via coupling to another pair state.
As a result, the pair state $\ket{r_1r_1}$ has vdW interaction $\sim C_6/R^{-6}$, which scales as $n^{11}$, and thus can be discretely tuned by changing the principal quantum number.

Here, we employ a microwave (MW) field with tunable Rabi frequency $\Omega_\mathrm{MW}$ and detuning $\Delta$, coupling $\ket{r_1}$ to a different parity state $\ket{r_2}$, and thus forming two dressed states $\ket{+}=\cos{\theta}\ket{r_1}+\sin{\theta}\ket{r_2}$ and $\ket{-}=\sin{\theta}\ket{r_1}-\cos{\theta}\ket{r_2}$, where $\tan{\theta}=-(\Delta+\sqrt{\Delta^2+\Omega_\mathrm{MW}^2})/\Omega_\mathrm{MW}$.
The relevant atomic levels are shown in figure~\ref{fig1}(a).
Since the dipole moment in $\ket{+}$ is no longer zero, the projection of $\bm{\hat{V_{dd}}}$ in the dressed pair state $\ket{++}$, i.e., the Rydberg-Rydberg interaction, becomes thus non-vanishing in the first order and dependent on the exact wave-function of $\ket{+}$. 
The engineering of dressed state $\ket{+}$ wave-function can be achieved by controlling the MW detuning $\Delta$, resulting in a transition of the Rydberg interaction from vdW for the bare state to dipole-dipole form $\sim C_3/R^{-3}$ for a maximally dressed state.

In Rydberg dephasing experiments, we are mainly interested in the interactions between atomic pairs with large inter-atomic distances $R$, since the pairs within the blockade radius are strongly suppressed.
Therefore, in the dressed-state basis, the non-diagonal elements of the dipole-dipole interaction operator $\bm{\hat{V_{dd}}}$ that couples the $\ket{++}$ state to other states can be treated perturbatively, leading to an effective Rydberg interaction $V^{++}$ for the dressed pair state $\ket{++}$ in the large-$R$ regime as (see Methods):
\begin{equation}
\begin{array}{rcl}
V^{++} &\approx& \frac{C_{3}^{++}}{R^{3}} + \frac{C_{6}^{++}}{R^{6}}, \\[6pt]
C_{3}^{++} &=& \frac{\mu_{12}}{2} \sin^{2}{2\theta}, \\[6pt]
C_{6}^{++} &=& \frac{\mu_{12}^{2} (\sin^{2}{4\theta} +2\sin^{4}{2\theta})}{16(E_{+}-E_{-})} 
+ \frac{\mu_{34}^{2} \cos^{4}{\theta}}{(2E_{+}-E_{3}-E_{4})}.
\end{array}
\end{equation}
where $C_{3}^{++}$ ($C_{6}^{++}$) is the dipole-dipole (vdW) interaction coefficient for the Rydberg dressed state.
$E_{+}$ ($E_{-}$) is the energy of the Rydberg dressed states $\ket{+}$ ($\ket{-}$), with $E_{3}$ and $E_{4}$ denoting the energy of other involved Rydberg states via dipole-dipole coupling.
The specific energy levels $E_3$ and $E_4$ depend on the experimental level scheme.
In our experimental setup, the rationale for employing the four-state model and the selection of energy levels $E_3$ and $E_4$ is detailed in Appendix B.
$\mu_{12}$ and $\mu_{34}$ represent the corresponding dipole moment.

\begin{figure*}[t]
  \centering
  \includegraphics[width=\textwidth]{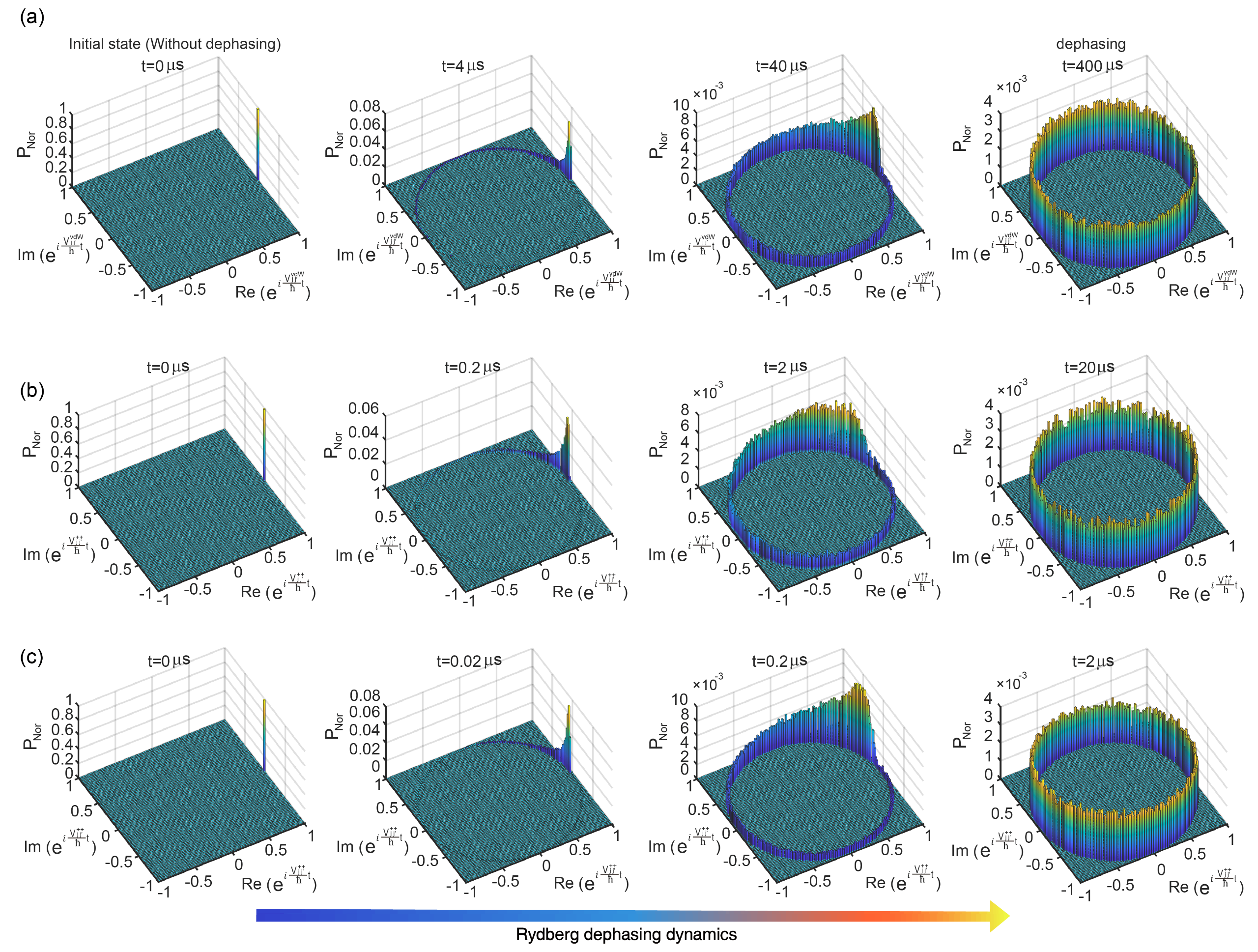}
  \caption{
    Simulation of Rydberg dephasing dynamics.
    (a) The exponential distribution $e^{iV^{vdW}_{jj\prime}t/\hbar}$ of the Rydberg interaction-induced phase at different interaction time.
    The x-axis represents the real part of the $e^{iV^{vdW}_{jj\prime}t/\hbar}$, while the y-axis represents its imaginary part.
    The z-axis corresponds to the normalized probability for a given phase.
    The result is simulated under the vdW interaction for bare state $\ket{r_1}=\ket{47D_{5/2}, J=5/2, m_J=5/2}$, via the Monte-Carlo method.
    (b) (c) Similar to (a) but displaying the exponential distribution  $e^{iV^{++}_{jj\prime}t/\hbar}$ for the Rydberg MW-dressed states with $\Delta=-2\Omega_\mathrm{MW}$ (b) and $\Delta=0$ (c).
    }
  \label{fig2}
\end{figure*}

Figure~\ref{fig1}(b) displays the interaction strength as a function of inter-atomic distances $R$ on a logarithmic scale. 
We first study the case of a bare-state pair $\ket{r_1r_1}$, where $\ket{r_1} = \ket{55D_{5/2}, J=5/2, m_J=5/2}$.
As shown in the blue curve, the bare-state pair exhibits the vdW interaction ($\sim R^{-6}$).
When the resonant MW coupling $\ket{r_1}$ to $\ket{r_2}=\ket{56P_{3/2}, J=3/2, m_J=3/2}$ is applied, the interaction is substantially enhanced to dipole-dipole form and scales as $\sim R^{-3}$ (lavender-colored curve).
By varying the MW detuning $\Delta$, the ratio of $\ket{r_2}$ in Rydberg dressed state $\ket{+}$ can be adjusted, and the transition of interaction from vdW interaction to dipole-dipole interaction or a mixture of both, can be achieved.
Therefore, for large distance $R \sim$ \SI{20}{\micro\meter}, the interaction strength can be continuously tuned by orders of magnitude.

Considering the scaling of dipole-dipole interaction as $n^4$, and vdW interaction as $n^{11}$, our microwave-dressed protocol demonstrates superior performance, particularly with low-lying Rydberg states.
We also study the interactions for lower $n$ states $\ket{r_1}=\ket{47D_{5/2}, J=5/2, m_J=5/2}$ and $\ket{r_2}=\ket{48P_{3/2}, J=3/2, m_J=3/2}$, displayed in figure~\ref{fig1}(c).
For example, when $R \sim$ \SI{20}{\micro\meter}, continuous tuning of interaction strength over nearly three orders of magnitude can be achieved by varying the dressing MW detuning $\Delta$.

\begin{figure}[t]
  \centering
  \includegraphics[width=\columnwidth]{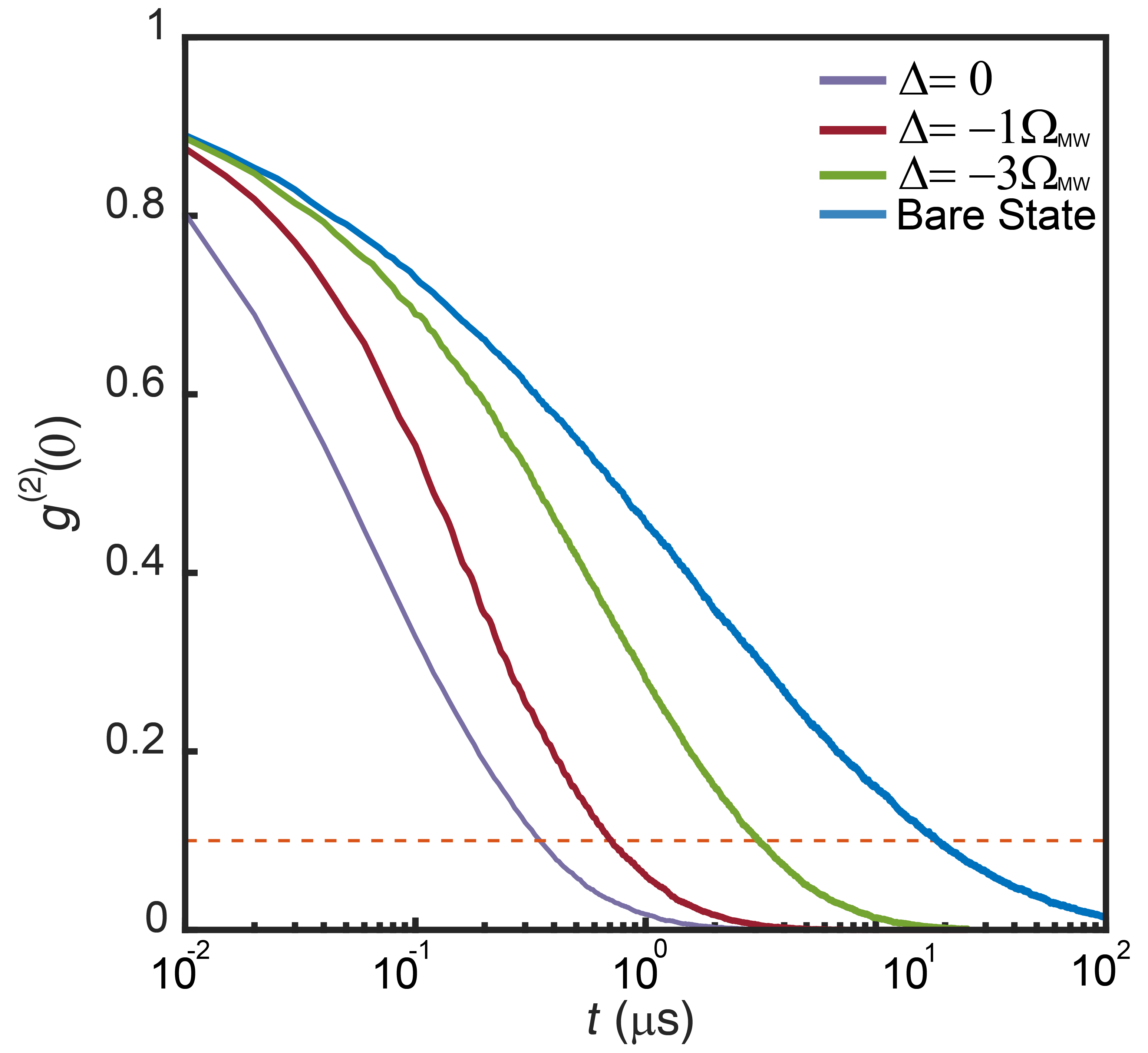}
  \caption{
    Acceleration of quantum operations via Rydberg-Rydberg dressing.
    Simulated $g^{(2)}(0)$ as a function of interaction time $t$ for atoms in the bare state $\ket{r_1}=\ket{47D_{5/2}, J=5/2, m_J=5/2}$, and dressed states.
    The orange-colored dashed line is the threshold for $g^{(2)}(0)=0.1$. 
    }
  \label{fig3}
\end{figure}

The dependence of $V^{++}$ on $\Omega_\mathrm{MW}$ and $\Delta$ enables the effective control of single-photon nonlinearity and the dephasing dynamics in Rydberg EIT photon storage.
Our Rydberg EIT scheme involves a ladder-type energy structure including a ground state $\ket{g}$, an intermediate state $\ket{e}$, and a MW-dressed Rydberg state $\ket{+}$.
Assisted by the \SI{480}{\nano\meter} control field $\Omega_{480}$ resonantly coupling the intermediate state $\ket{e}$ to the MW-dressed Rydberg state $\ket{+}$, the input \SI{780}{\nano\meter} photons are stored as collective Rydberg excitations.
It's clear from figure~\ref{fig1}(b),(c) that $V^{++}_{jj\prime}$, the interaction between two Rydberg-dressed atoms $\ket{+}_{j}$ and $\ket{+}_{j\prime}$, strongly depends on the atomic pair separations $R_{jj\prime}$.
Therefore, atomic pairs with different separations $R_{jj\prime}$ in stored Rydberg excitations interact differently and accumulate random phases during the storage time $t$, leading to collective Rydberg excitations dephasing \cite{bariani2012dephasing}. This dephasing process can be illustrated by the time evolution of a doubly-excited dressed state $\ket{++}$:
\begin{equation}
    \ket{++}\propto
    \sum^N_{j,j^\prime \neq j}{
    e^{i\frac{V^{++}_{jj\prime}}{\hbar}t}
    \ket{\mathrm{g}}_1 \ldots
    \ket{\mathrm{+}}_j \ldots
    \ket{\mathrm{+}}_{j^\prime} \ldots
    \ket{\mathrm{g}}_N
    }.
\end{equation}
This distance-dependent quantum evolution leads to the accumulation of random phases and deteriorates the phase-matching condition over time.
To illustrate this interaction-induced dephasing dynamics, we employ Monte-Carlo simulation to obtain the distribution of the Rydberg interaction-induced phase term $e^{iV^{++}_{jj\prime}t/\hbar}$ with different evolution times.
The numerical simulation is conducted based on the two-body decoherence model \cite{bariani2012dephasing, stanojevic2012generating, busche2017contactless, ye2023photonic}, considering the coupling of Rydberg states $\ket{r_1} = \ket{47D_{5/2}, J=5/2, m_J=5/2}$ and $\ket{r_2} = \ket{48P_{3/2}, J=3/2, m_J=3/2}$.
The excitation region involves an atomic ensemble with a uniform distribution spanning \SI{9}{\micro\meter} along the x and y directions and a Gaussian distribution with a $1/e^2$ parameter of \SI{3.5}{\micro\meter} along the z direction, containing totally $440$ atoms corresponding the atomic density of $\sim 5 \times 10^{11}$ \SI{}{\centi\meter^{-3}}.
The influence of the Gaussian intensity distribution of the excitation beam is subsequently considered through a weighted excitation probability during the EIT photon storage process.
All these parameters are chosen to closely align with our experimental configurations.

\begin{figure}[t]
  \centering
  \includegraphics[width=\columnwidth]{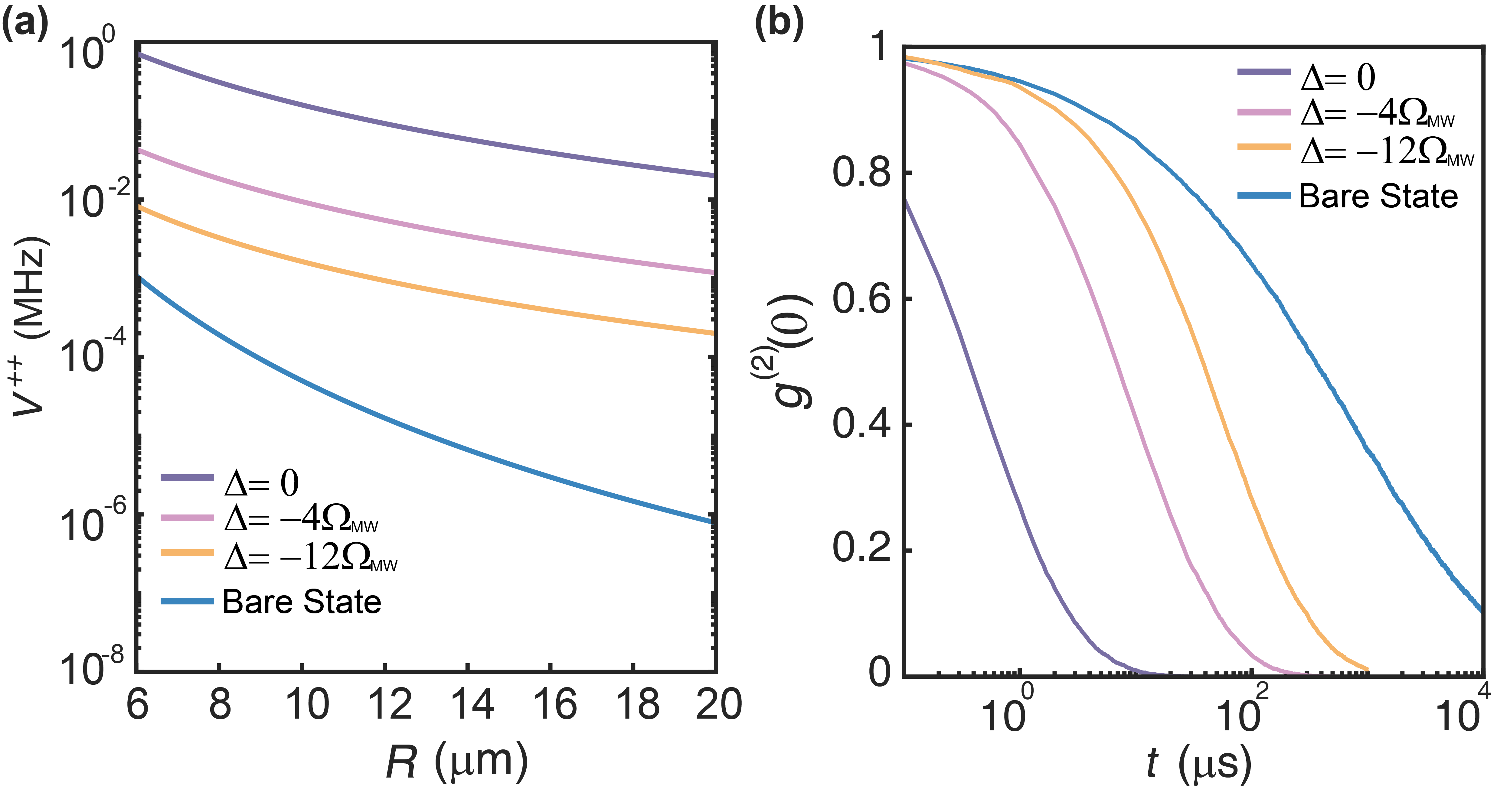}
  \caption{
    Enhanced interactions and dephasing dynamics with low-lying Rydberg states.
    (a) The interaction strength on a logarithmic scale as a function of inter-atomic distance $R$ for low-lying Rydberg state $\ket{r_1}=\ket{29D_{5/2}, J=5/2, m_J=5/2}$ and dressed states with different MW detunings.
    (b) Simulated $g^{(2)}(0)$ as a function of interaction time $t$ for the bare state and for the MW-dressed Rydberg states under the interactions shown in (a).
    }
  \label{fig4}
\end{figure}

As shown in figure~\ref{fig2}, all the Rydberg atom pairs have the same initial phase evolution term $e^{iV^{++}_{jj\prime}t/\hbar}=1$ at t=0.
If photon read-out is performed immediately, that is, without two-body interaction-induced dephasing, the spatial mode of the retrieved photons is highly directional due to the phase-matching collective emission \cite{lukin2001dipole, saffman2002creating}.
However, as interaction time $t$ increases, the Rydberg excitations accumulate random phases due to the distance-dependent interaction $V^{++}_{jj\prime}$, leading to a broadened distribution of $e^{iV^{++}_{jj\prime}t/\hbar}$ [figure~\ref{fig2}]. 
Notably, the evolution of the distribution of the interaction-induced phase strongly depends on the strength of the Rydberg interactions.
As shown in figure~\ref{fig2}(b), when the MW field with $\Delta=-2\Omega_\mathrm{MW}$ is applied, the interaction between Rydberg dressed states $\ket{+}$ is significantly enhanced, resulting in a faster deterioration of $e^{iV^{++}_{jj\prime}t/\hbar}$ compared to the bare state [figure~\ref{fig2}(a)].
If resonant MW-dressing is applied, the Rydberg interaction can be further improved, accelerating the Rydberg dephasing dynamics by orders of magnitude [figure~\ref{fig2}(c)].

Such an interaction dephasing mechanism would largely change the quantum statistics of coherent light pulses because of the Rydberg interaction.
The nonlinearity is induced under the distance-dependent interaction during the interaction time, which smears the collective phase coherence.
As a result, in the photon readout process, the multi-photon components will be emitted in random directions over the $4\pi$ solid angle instead of exhibiting highly directional emission as the single-photon component.
Therefore, the second-order correlation function $g^{(2)}(0)$ of coherent light would decrease after transmitting the atoms, and the fast decay of $g^{(2)}(0)$ will characterize the higher nonlinearity.

\begin{figure*}[t]
  \centering
  \includegraphics[width=\textwidth]{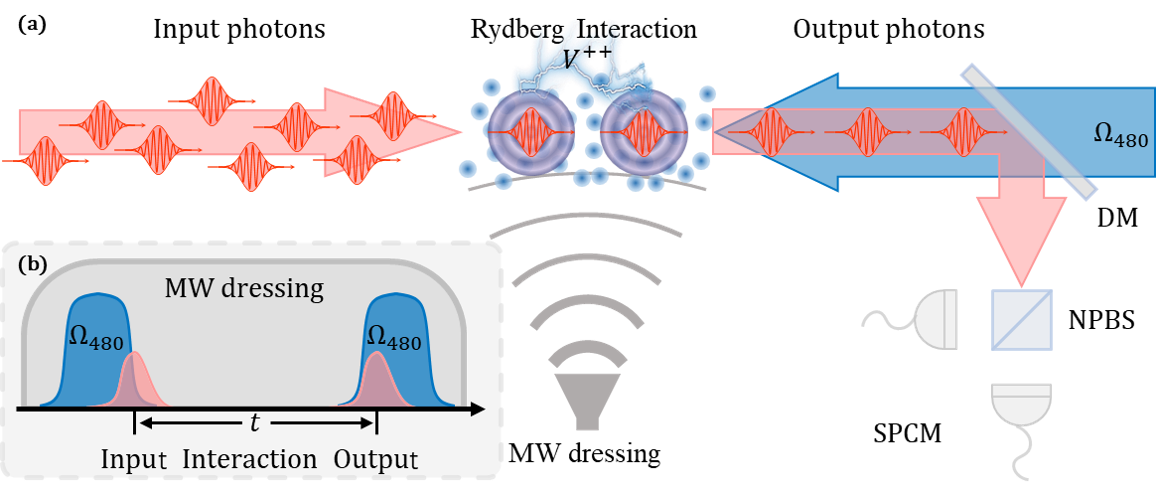}
  \caption{
     Illustration of the experimental protocol.
     (a) A cold $^{87}$Rb atomic ensemble with a temperature of $\sim$ \SI{10}{\micro\kelvin} and an optical depth of $\sim 3.5$ is confined in a \SI{1012}{\nano\meter} optical dipole trap.
     The counter-propagating \SI{780}{\nano\meter} and \SI{480}{\nano\meter} laser beams are focused onto the atomic ensemble.
     The quantum statistics of the retrieved photons are measured using a Hanbury Brown-Twiss (HBT) setup consisting of a 50:50 beam splitter, followed by two single-photon counting modules (SPCMs).
     (b) Experimental timing sequence.
     The total duration of the storage-and-retrieval process is \SI{5}{\micro\second} covered by the dressing microwave, shown as the gray wave packet.
     The \SI{780}{\nano\meter} input field is a $\SI{50}{\nano\second}$ long weak coherent light pulse with a mean photon number $\braket{n}=0.4$, represented by the pink wave packet.
     By optimizing the falling edges of the blue \SI{480}{\nano\meter} control field to maximize the storage efficiency ($24\%$ limited by the optical depth OD $\sim 3.5$), the input photons are stored in the atomic ensemble.
     The interaction time $t$ in the experiment is variable, ranging from \SI{190}{\nano\second} to \SI{1000}{\nano\second} in the experiment, and can be extended to its decoherence time in principle.}
  \label{fig5}
\end{figure*}

Since the Rydberg dephasing dynamics can be expedited by our MW dressing protocol, one would expect it can also speed up the dephasing-based single photon preparation process.
Numerical simulations are conducted to obtain quantum statistics of retrieved photons as a function of storage time $t$, assuming coherent light pulse with mean photon number $\braket{n}=0.4$ is stored into the atoms via Rydberg EIT.
Figure~\ref{fig3} shows evolution of the second-order intensity correlation function $g^{(2)}(0)$ with different interaction strength.
If we set $g^{(2)}(0)<0.1$ as the criteria for good single photons, it would take at least \SI{18}{\micro\second} evolution time in the bare state scenario (blue curve).
Achieving such a long coherence time between the atomic ground and the Rydberg state is challenging, and protocols such as the Rydberg-ground magic lattice \cite{lampen2018long, mei2022trapped} might be needed.
In contrast, with the application of our Rydberg-Rydberg dressing protocol, the enhanced interaction results in a much faster evolution of $g^{(2)}(0)$. 
Our simulation suggests that at approximately \SI{400}{\nano\second}, the value of $g^{(2)}(0)$ would reach $g^{(2)}(0)<0.1$, accelerated by a factor of $\sim 40$ times compared to the bare state case.

Many advanced photonic quantum information applications post even higher requirements on the purity ($g^{(2)}(0)$) of the single photons.
For example, $g^{(2)}(0)<0.01$ is required for single-photon sources used in memory-based quantum repeaters~\cite{aharonovich2016solid}.
Such high-purity single photons are almost unattainable via the conventional dephasing protocol with low Rydberg states.
We simulated it would take an interaction time of \SI{124}{\micro\second} to achieve $g^{(2)}(0)<0.01$ using Rydberg state $\ket{r_1}=\ket{47D_{5/2}, J=5/2, m_J=5/2}$, much longer than its $\sim$ \SI{58}{\micro\second} lifetime.
However, if our MW-dressing technique is applied, $g^{(2)}(0)<0.01$ can be achieved within just \SI{1.3}{\micro\second}.
Our simulated results indicate Rydberg wave-function dressing holds the potential to realize highly demanding photonic quantum operations using low-lying Rydberg states.

\begin{figure*}[t]
  \centering
  \includegraphics[width=\textwidth]{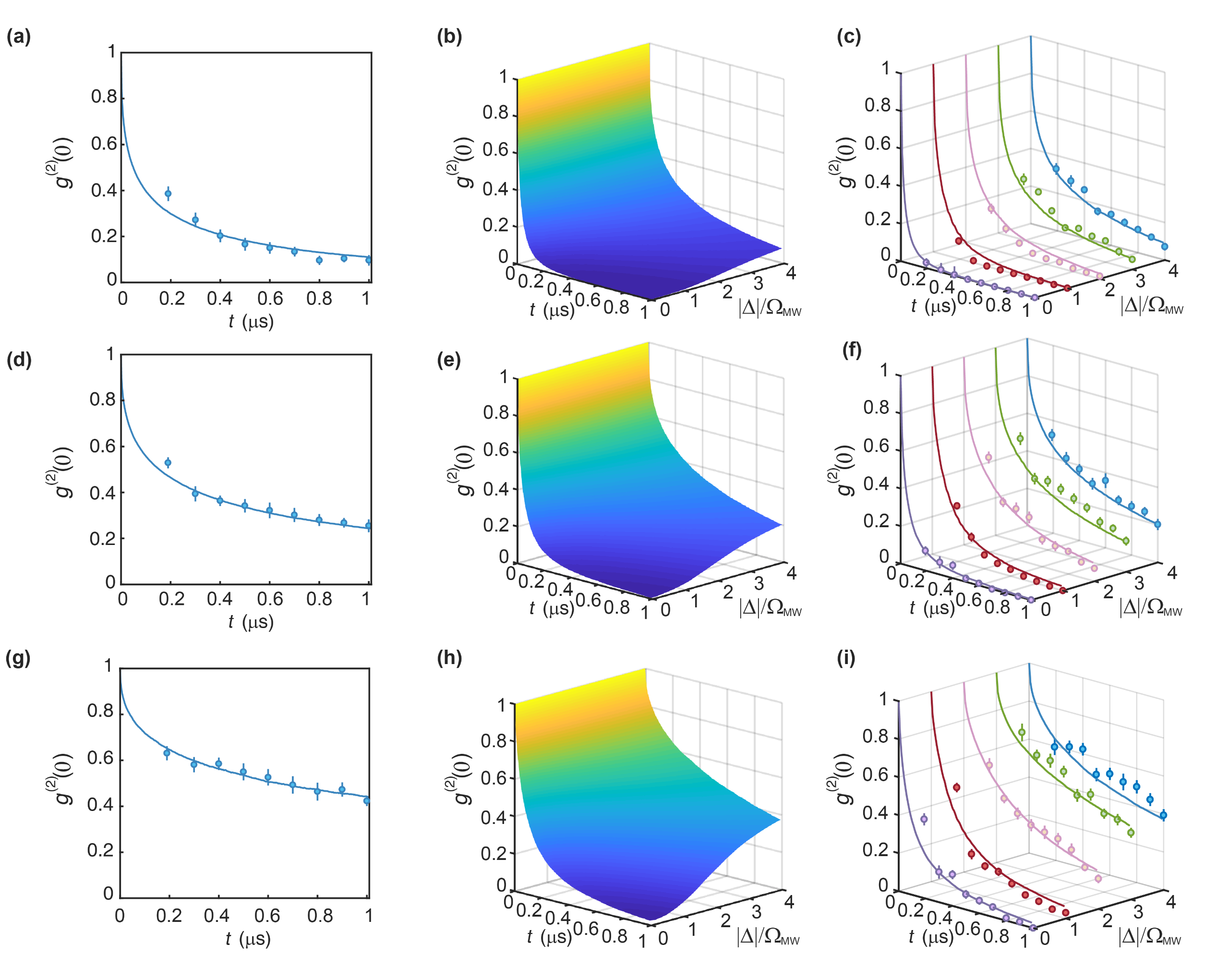}
  \caption{
    Dephasing dynamics with continuously tunable Rydberg interactions.
    (a) (c) Measured second-order intensity correlation function $g^{(2)}(0)$ when the MW field is off (a) and on (c).
    Solid lines are theoretical simulations.
    (b) is the simulated $g^{(2)}(0)$ dynamics under different detuning $\Delta$.
    $\ket{r_1}=\ket{55D_{5/2}, J=5/2, m_J=5/2}$ and $\ket{r_2}=\ket{56P_{3/2}, J=3/2, m_J=3/2}$ are employed.
    The waist of the \SI{780}{\nano\meter} laser beam is \SI{6.5}{\mu \meter}.
    (d-f) same as (a-c), but with a \SI{9}{\mu \meter}-waist \SI{780}{\nano\meter} excitation beam.
    (g-i) same as (d-f), but with a lower principal quantum number $n=47$ and the state $\ket{r_1}=\ket{47D_{5/2}, J=5/2, m_J=5/2}$ is coupled to $\ket{r_2}=\ket{48P_{3/2}, J=3/2, m_J=3/2}$ via the MW field.
    Error bars: $1\sigma$ standard deviation from photoelectric counting events.
    }
  \label{fig6}
\end{figure*}

The MW-dressing-induced interaction enhancement and accelerated dephasing process become even more prominent with lower principal quantum numbers $n$.
To further illustrate this, we consider the state  $\ket{r_1}=\ket{29D_{5/2}, J=5/2, m_J=5/2}$, whose vdW interaction is less than \SI{}{\kilo\hertz} for $\sim$ \SI{10}{\micro\meter} atomic distance [blue curve in figure~\ref{fig4}(a)].
Therefore, interaction dephasing and the decrease of $g^{(2)}(0)$ happen on a very long timescale [blue curve in figure~\ref{fig4}(b)].
This weak nonlinearity can be drastically enhanced by coupling $\ket{r_1}$ to a state $\ket{r_2}=\ket{30P_{3/2}, J=3/2, m_J=3/2}$, which is experimentally accessible by employing a $\sim$ \SI{94}{\giga\hertz} MW dressing field.
Figure~\ref{fig4}(a) depicts the interaction strength as a function of inter-atomic distance $R$ for the MW-dressed Rydberg states with different detunings $\Delta$, showing 3 to 4 orders of magnitude interaction enhancement compared to the bare state.
The corresponding dephasing dynamics are presented in figure~\ref{fig4}(b), providing an evident demonstration of the acceleration for the quantum operation through the MW-dressed protocol.
Again, we take the generation of high-quality single photons ($g^{(2)}(0)<0.1$) as an example.
For bare Rydberg states, achieving such high-quality single photons is impossible since it requires an extremely long interaction time (\SI{10}{\milli\second}), excessively surpassing the lifetime of the Rydberg state $\ket{r_1}$.
In striking contrast, with the resonant MW dressing technique, \SI{2.5}{\micro\second} interaction time is enough to generate these high-quality single photons, leading to a remarkable speed-up factor of $\sim 4000$.

\section{Experimental protocol and result}
\noindent
Next, we experimentally demonstrate the continuous tuning of single-photon level nonlinearity mediated by low-lying microwave-dressed Rydberg states. 
Our experiment is illustrated in figure~\ref{fig5}.
An ensemble of cold $^{87}$Rb atoms is employed as the nonlinear quantum medium for photon-photon interaction control.
The atoms are first collected in a magneto-optical-trap from the background vapor and then transferred to a one-dimensional optical dipole trap.
The optical dipole trap operates at \SI{1012}{\nano\meter} with a power of \SI{0.9}{\watt} and an elliptical cross-section (waist sizes of \SI{10}{\micro\meter} and \SI{50}{\micro\meter}), resulting in a trap depth of $U/h \sim$ \SI{4}{\mega\hertz}.
The $^{87}$Rb sample is cooled to \SI{10}{\micro\kelvin} by polarization-gradient cooling.
An external magnetic field of \SI{8}{G} is applied, and the atoms are optically pumped to the $\ket{g}=\ket{5S_{1/2}, F=2, m_F=2}$ state.

Both the \SI{480}{\nano\meter} and \SI{780}{\nano\meter} laser fields have $\sigma^+$  polarization, while the $\sigma^-$  polarized component of the MW field is used.
The \SI{780}{\nano\meter} and \SI{480}{\nano\meter} control fields are tightly focused on the ensemble, with waists of \SI{6.5}{\micro\meter} and \SI{17}{\micro\meter}, respectively.
The \SI{780}{\nano\meter} input field is a \SI{50}{\nano\second} long weak coherent light pulse with the mean photon number $\braket{n}=0.4$.
A MW field couples the Rydberg state $\ket{r_1}=\ket{nD_{5/2}, J=5/2, m_J=5/2}$ to an adjacent one $\ket{r_2}=\ket{(n+1)P_{3/2}, J=3/2, m_J=3/2}$ with a detuning $\Delta$ and Rabi frequency $\Omega_\mathrm{MW} \sim$ \SI{12}{\mega\hertz}.
By properly selecting $\Omega_\mathrm{MW}$ and detuning of the \SI{480}{\nano\meter} control field, the \SI{780}{\nano\meter} input photons can be stored in the Rydberg dressed state $\ket{+}$ via the Rydberg EIT photon storage technique.
The size of the cold atomic ensemble and \SI{780}{\nano\meter} field are chosen such that the photon storage region is larger than the blockade radius.
Therefore, the multi-photon components from the input coherent light can be stored into the atomic ensemble.
After the photon storage, the Rydberg excitations interact and accumulate random phases, leading to the dephasing of collective atomic excitation.
After an interaction time $t$, the Rydberg excitations are converted back to photons through collective emission by applying again the \SI{480}{\nano\meter} read-out light field.
The measured storage efficiency and read-out efficiency are $24\%$ and $36\%$, respectively, which are currently limited by the finite optical depth OD $\sim 3.5$. 
At the end of each trial, the resonant \SI{480}{\nano\meter} field is turned on for \SI{2}{\micro\second} to eliminate any remaining Rydberg contaminants that could affect subsequent Rydberg EIT storage.
The experimental timing sequence for the storage-and-retrieval process lasts \SI{5}{\micro\second}, with a repetition rate of $15,000$ times per second.

To characterize the single-photon nonlinearity and study the Rydberg superatom's dephasing dynamics, we measure the quantum statistics of the output photons.
The output photons along phase-matched direction are collected into a 50:50 fiber beam splitter and detected by two single-photon counting modules (SPCMs).
Figure~\ref{fig6}(a) shows the measured second-order intensity correlation function $g^{(2)}(0)$ without the MW dressing field.
As the storage time $t$ increases from \SI{190}{\nano \second} to \SI{1000}{\nano\second}, $g^{(2)}(0)$ of the bare state $\ket{r_1}$ drops from $0.386(32)$ to $0.096(23)$, which agrees well with our simulated curve [solid line in figure\ref{fig6}(a)] based on the two-body decoherence model \cite{bariani2012dephasing, stanojevic2012generating}.
This dephasing dynamics for the bare state is dominated by the vdW interaction of Rydberg atoms in state $\ket{r_1}=\ket{55D_{5/2}, J=5/2, m_J=5/2}$. 

When the MW dressing protocol is applied, with the MW-frequency of \SI{12.704}{\giga\hertz}, the coefficient of $\ket{r_1}$ and $\ket{r_2}$ components in the dressed states $\ket{+}$ change with different detuning $\Delta$, leading to a $\Delta$-dependent Rydberg interaction and single-photon nonlinearity.
Using the aforementioned two-body decoherence model, the dephasing dynamics under different $\Delta$ is simulated and shown in figure~\ref{fig6}(b).
It is obvious that the MW dressing leads to stronger nonlinearity and thus faster dephasing.
To experimentally verify the simulated results, we measure the time evolution of $g^{(2)}(0)$ with $\Delta=0, -1, -2, -3$ and $-4 \Omega_\mathrm{MW}$ [figure~\ref{fig6}(c)].
In the resonant case ($\Delta=0$), $g^{(2)}(0)$ dramatically dropped to $5(5) \times 10^{-3}$ within just $t=$ \SI{400}{\nano \second}, while the $g^{(2)}(0)$ for $t=$ \SI{400}{\nano \second} in the bare-state case is still high ($0.203(28)$).
Although the value obtained in the resonant case is an order of magnitude smaller than the $g^{(2)}(0)=0.1$ threshold, we can further improve the purity of this single-photon source by extending its interaction time, potentially up to the coherence time.
As $\Delta$ increases, although the decay of $g^{(2)}(0)$ is slower, the dephasing dynamics still happen much quicker than the bare-state because of the enhanced single-photon nonlinearity.

Our Rydberg wave-function engineering protocol is capable of not only controlling the interaction strength but also its dependence on the inter-atomic separation $R_{jj\prime}$.
For example, the resonant dressing changes the interaction from vdW  ($\sim R^{-6}$ scaling) to dipole-dipole ($\sim R^{-3}$ scaling).
Therefore, for large atomic ensembles, the advantage of our protocol over the bare-sate case becomes very obvious.
To study this effect, we enlarge the root-mean-squared (RMS) separation between the Rydberg atoms by increasing the waist of the \SI{780}{\nano\meter} field from \SI{6.5}{\micro\meter} to \SI{9}{\micro\meter}.
The decreased vdW interaction slows down the dephasing dynamics for the bare state [figure~\ref{fig6}(d)].
Even with \SI{1}{\micro \second} interaction time, $g^{(2)}(0)$ is still $0.253(28)$.
Figures~\ref{fig6}(e)-(f) display the dephasing dynamics for the simulated and measured MW dressed state and clearly demonstrate the acceleration of quantum operation.
These results suggest our protocol holds the promise for fast single-photon generation in a large Rydberg ensemble.

\begin{figure}[t]
  \centering
  \includegraphics[width=0.95\columnwidth,height=0.95\columnwidth]{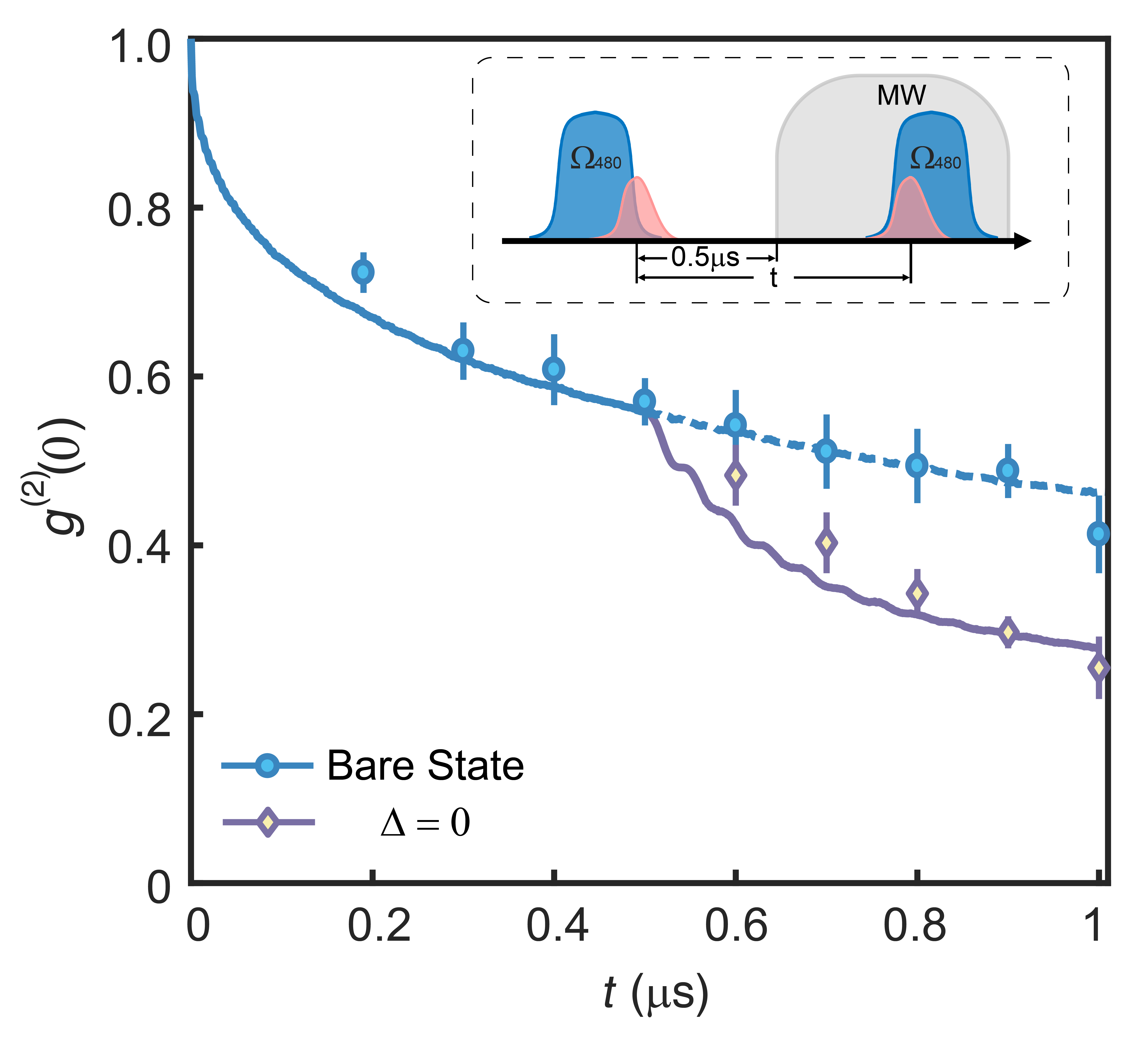}
  \caption{
    Dynamical control of the Rydberg dephasing dynamics.
    Measured $g^{(2)}(0)$ as a function of the storage time $t$ with principal quantum number $n=47$ and \SI{9}{\micro\meter}-waist \SI{780}{\nano\meter} excitation beam.
    The blue solid and dashed lines represent simulations of Rydberg dephasing dynamics dominated by the vdW interaction of the bare state, while the lavender-colored curve shows the simulated result when the resonant MW field is applied.
    The error bars represent the $1\sigma$ standard deviation from photoelectric counting events.
    Inset: timing sequence for dynamical control of the dephasing dynamics.
    }
  \label{fig7}
\end{figure}

Our protocol also features advantages over bare-state in the case of low-lying Rydberg states.
We further reduce the Rydberg dephasing rate by choosing a lower principal quantum number $n=47$ while keeping the waist of \SI{780}{\nano\meter} field as \SI{9}{\micro\meter}. 
As shown in figure~\ref{fig6}(g), the measured second-order intensity correlation function $g^{(2)}(0)$ for the bare state decreases slowly from $0.631(31)$ to $0.423(23)$ as the interaction time $t$ increases from \SI{190}{\nano\second} to \SI{1000}{\nano\second}.
Our simulation suggests that preparing good single photons with $g^{(2)}(0) < 0.1$ using such a low $n$ state would take more than $\SI{18}{\micro\second}$ of interaction time, which poses stringent requirements on the ground-Rydberg level coherence times.
Coherence times on the order of tens \SI{}{\micro\second} have been achieved via the simultaneous confinement of ground and Rydberg atoms in a state-insensitive optical lattice \cite{lampen2018long, mei2022trapped, li2022dynamics}.
Alternatively, we can employ the MW dressing protocol to enhance the interaction and accomplish the operation within a much shorter time.
Here, the state $\ket{r_1}=\ket{47D_{5/2}, J=5/2, m_J=5/2}$ that is MW dressed with $\ket{r_2}=\ket{48P_{3/2}, J=3/2, m_J=3/2}$ by the \SI{20.653}{\giga\hertz} MW, and the corresponding simulated and measured dephasing dynamics are shown in figure~\ref{fig6}(h), (i).
Within \SI{500}{\nano\second}, a single photon with $g^{(2)}(0) = 0.082(23)$ is achieved with the enhanced interaction, offering a factor of $\sim 40$ acceleration compared to the bare-state case.

The above results demonstrate the ability to manipulate single-photon nonlinearity and dephasing dynamics with Rydberg state wave-function dressing.
The interaction time $t$ and detuning $\Delta$ can be varied continuously, and therefore, we can achieve the continuous two-dimensional engineering of photon quantum statistics $g^{(2)}(0)$.
Even when considering its practicability as a single-photon source, our experimental result exhibits a state-of-the-art level of purity for Rydberg-dephasing-based single-photon sources.
In principle, the microwave-dressed dephasing scheme also works remarkably well with even lower principal quantum numbers.
As we discussed in section II, for low-lying states (e.g., $n \sim 29$), our protocol could provide orders of magnitude interaction enhancement and quantum operations acceleration.

An additional advantage of our MW-dressed scheme is its capability to dynamically manipulate the strength of single-photon nonlinearity even within a single experimental sequence, which is unattainable through the conventional methods of altering the principal quantum number.
To demonstrate this capability, we employ the Rydberg dressing protocol to dynamically control dephasing dynamics in the photon storage and retrieval process.
The dynamic control is implemented by selectively turning on the MW field during a single experimental cycle, with the timing sequence shown as the insert in figure~\ref{fig7}.
During first \SI{0.5}{\micro\second} evolution, the MW field remains switched off, resulting in the system undergoing dephasing dynamics dominated by the vdW interaction of the bare state $\ket{r_1}=\ket{47D_{5/2}, J=5/2, m_J=5/2}$.
As the storage time $t$ increases from \SI{190}{\nano\second} to \SI{500}{\nano\second}, the measured $g^{(2)}(0)$ gradually decreases from $0.723(24)$ to $0.570(28)$, in agreement with our simulated results (blue solid curve).

If the microwave field remains off during the rest of the experiment, the system continues to experience dephasing dynamics governed by the vdW interaction, following the simulated trajectory indicated by the blue dashed curve.
As a result,  $g^{(2)}(0)$ gradually decays to $0.413(46)$ when the storage time $t$ reaches \SI{1000}{\nano\second}.
In contrast, when the resonant MW field is turned on during the subsequent \SI{0.5}{\micro\second}, the coupling between the Rydberg states $\ket{r_1}$ and $\ket{r_2}$ leads to enhanced Rydberg interactions, and consequently, a stronger nonlinearity strength.
In this case, a faster Rydberg dephasing dynamics, evolving along the simulated trajectory indicated by the lavender-colored solid line, has been observed.
$g^{(2)}(0)$ decays to 0.225(23) at $t=$ \SI{1000}{\nano\second}.

\section{Summary and outlook}

\noindent
In summary, we develop a new scheme for the continuous control of photon-photon interaction by manipulating the Rydberg-Rydberg wave-function dressing.
We demonstrate the tunability of Rydberg dephasing dynamics and moreover, show the ability to dynamically control the dephasing process.
By combining the two techniques, continuous engineering for single-photon level nonlinearity can be achieved.
Our scheme enables fast preparation of high-quality single photons using large Rydberg ensembles and low principal quantum number $n$.
The results demonstrated here open new avenues for novel quantum photonic applications based on low-lying Rydberg states.
Compared with the blockade protocol using high $n$ states, dephasing-based scheme with low $n$ states is less prone to various decoherence and losses such as long-lived Rydberg contaminants, energy level shifts induced by residual electric fields, and Rydberg-ground interaction decoherence.

We emphasize that although our interaction control protocol is used to speed up the dephasing-based single photon generation, it is, in principle, applicable to many other Rydberg quantum operations, such as single photon transistor, entanglement filter, and entangled photon creation. 
In particular, our Rydberg-Rydberg dressing protocol holds the promise to improve the circuit depth of Rydberg-based quantum computing architectures.
Leading quantum computing platforms, such as trapped ions or superconducting qubits, have circuit depth up to a few hundred \cite{arute2019quantum, cao2023generation, kim2023evidence, kim2023scalable}, which is challenging to achieve in current Rydberg atoms experiments, due to the finite coherence times, lifetimes, and quantum operation speed.
Our dressed state protocol can be used in Rydberg-interaction-enabled quantum logic gates between photon-photon, atom-atom, and atom-photon qubits, with the potential to accelerate the operation time and increase the circuit depth by orders of magnitude.
It can also improve the connectivity in a Rydberg quantum computer, since stronger interaction leads to a larger blockade radius.
Moreover, the capability to tune the interaction strength efficiently in a broad energy scale enables us to explore novel non-equilibrium states in many-body systems \cite{martinez2019tunable}, such as the dynamical formation of the Wigner crystal \cite{otterbach2013wigner} and the Kibble-Zurek physics across a quantum phase transition \cite{clark2020observation, li2021symmetry, sahay2021emergent}.

\section*{Acknowledgments}
We thank Daiqin Su for valuable discussions and Shuai Shi and Feng-Yuan Kuang for experimental assistance.
We further thank the open-source library `ARC'. 
This work was supported by the National Key Research and Development Program of China under Grant No.~2021YFA1402003, the National Natural Science Foundation of China (Grant No.~12374329 and No.~U21A6006.), and the Fundamental Research Funds for the Central Universities, HUST.
Y.C. is supported by the National Natural Science Foundation of China (Grant No.~U2141237).
T.S. is supported by National Key Research and Development Program of China (Grant No.~2017YFA0718304), and the National Natural Science Foundation of China (Grants No.~11974363, No.~12135018, and No.~12047503).

\section*{Data acailabilty statement}
Data that support the plots within this paper are available from the corresponding authors upon reasonable request.

\section*{Conflict of interests}
The authors declare no competing interests.

\section*{References}

\providecommand{\newblock}{}

\end{document}